\documentclass[12pt,epsf,epsfig]{article}
\usepackage{amsmath}

\begin{document}

\begin{center}
{\huge{ Light Like Segment Compactification and Braneworlds with Dynamical  String Tension }}  \\
\end{center}

\begin{center}
 E.I. Guendelman  \\
\end{center}

\begin{center}
\ Department of Physics, Ben-Gurion University of the Negev, Beer-Sheva, Israel \\
\end{center}

\begin{center}
\ Frankfurt Institute for Advanced Studies, Giersch Science Center, Campus Riedberg, Frankfurt am Main, Germany \\
\end{center}

\begin{center}
\ Bahamas Advanced Studies Institute and Conferences,  4A Ocean Heights, Hill View Circle, Stella Maris, Long Island, The Bahamas \\
\end{center}
E-mail:  guendel@bgu.ac.il,     

\abstract
There is great interest in the construction of brane worlds, where matter and gravity are forced to be effective only in a lower dimensional surface , the ¨brane¨ . How these could appear as a consequence of string theory is a crucial question and this has been widely discussed. Here we will examine a distinct scenario that appears in dynamical string tension theories and where string tension is positive between two surfaces separated by a short distance and  at the two surfaces themselves the string tensions become infinite, therefore producing an effective confinement of the strings and therefore of all matter and gravity to the space between these to surfaces, which is in fact a new type of stringy brane world scenario.
The specific model studied is in the context of the modified measure formulation the string where tension appear as an additional dynamical degree of freedom and these tensions are not universal, but rather each string  generates its own  tension, which can have a different value for each string.  We consider a new background field that can couple to these strings,  the ¨tension scalar¨ is capable then of changing locally along the world sheet and then the value of the tension of the extended object changes accordingly. When many types of strings probing the same region of space are considered this tension scalar is constrained by the requirement of quantum conformal invariance. For the case of two types of strings probing the same region of space with different dynamically generated tensions, there are two different metrics, associated to the different strings,  that have to satisfy vacuum Einsteins equations and the consistency of these two Einstein´s equations determine the tension scalar. The universal metric, common to both strings generically does not satisfy Einstein´s equation . The two metrics considered here are flat space in Minkowshi space and flat space after a special conformal transformation and the tension field behaves in such a way that strings are confined inside a light like Segment or alternatively as expanding Braneworlds where the strings are confined between two expanding bubbles separated by a very small distance at large times.

\section{Introduction}

The basic idea of the brane worlds is that the universe is restricted to a brane inside a higher-dimensional space, called the "bulk" . In this model, at least some of the extra dimensions are extensive (possibly infinite), and other branes may be moving through this bulk. 
Some of the first braneworld models were developed by Rubakov and Shaposhnikov \cite{Rubakov}, Visser \cite{Visser}, Randall and Sundrum \cite{Randall1}, \cite{Randall2}, Pavsic \cite{Matej}, Gogberashvili \cite{Gogberashvili}. At least some of these models are motivated by string theory.  Braneworlds in string theory were discussed in \cite{Antoniadis}, see for a review for example \cite{Dieter Lust},
our approach will be very different to the present standard approaches to braneworlds in the context of string theories however. In our approach a dynamical string tension is required. Our scenario could be enriched by incorporating aspects of the more traditional braneworlds, but these aspects will be ignored here to simplify the discussion.

String  Theories have been considered by many physicists for some time as the leading candidate for the theory everything,  including gravity, the explanation of all the known particles that we know and all of their known interactions (and probably more) \cite{stringtheory}. According to some, one unpleasant feature of string theory as usually formulated is that it has a dimension full parameter, in fact, its fundamental parameter , which is the tension of the string. This is when formulated the most familiar way.
The consideration of the string tension as a dynamical variable, using the modified measures formalism, which was  previously used for a certain class of modified gravity theories under the name of Two Measures Theories or Non Riemannian Measures Theories, see for example \cite{d,b, Hehl, GKatz, DE, MODDM, Cordero, Hidden}. In the contex of this paper, it is also interesting to mention that the modified measure approach has also been used to construct braneworld scenarios 
\cite{modified measures branes}

When applying these principles to string theory, this leads to the  modified measure approach to string theory, where  rather than to put the string tension by hand it appears dynamically.

This approach has been studied in various previous works  \cite{a,c,supermod, cnish, T1, T2, T3, cosmologyandwarped}. See also the treatment by Townsend and collaborators for dynamical string tension \cite{xx,xxx}.

In our most recent papers on the subject \cite{cosmologyandwarped},
we have also introduced the ¨tension scalar¨, which is an additional
background
fields that can be introduced into the theory for the bosonic case (and expected to be well defined for all types of superstrings as well) that changes the value of the tension of the extended object along its world sheet, we call this the tension scalar for obvious reasons. Before studying issues that are very special of this paper we review some of the material contained in previous papers,  first present the string theory with a modified measure and containing also gauge fields that like in the world sheet, the integration of the equation of motion of these gauge fields gives rise to a dynamically generated string tension, this string tension may differ from one string to the other.

Then we consider the coupling of gauge fields in the string world sheet to currents in this world sheet, as a consequence this coupling induces variations of the tension along the world sheet of the string. Then we consider a bulk scalar and how this scalar naturally can induce this world sheet current that couples to the internal gauge fields. The integration of the equation of motion of the internal gauge field lead to the remarkably simple equation that the local value of the tension along the string is given by $T= e \phi + T _{i} $ , where $e$ is a coupling constant that defines the coupling of the bulk scalar to the world sheet gauge fields and  $ T _{i} $ is an integration constant which can be different for each string in the universe. 

Then each string is considered as an independent system that can be quantized. We take into account the string generation by introducing the tension as a function of the scalar field as a factor inside a Polyakov type action with such string tension, then the metric and the factor $g \phi + T _{i} $  enter together in this effective action, so if there was just one string the factor could be incorporated into the metric and the condition of world sheet conformal invariance will not say very much about the scalar  $\phi $ , but if many strings are probing the same regions of space time, then considering a background metric $g_{\mu \nu}$ , for each string the ¨string dependent metric¨  $(\phi + T _{i})g_{\mu \nu}$ appears and in the absence of othe background fields, like dilaton and antisymmetric tensor fields, Einstein´s equations apply for each of the metrics 
$(\phi + T _{i})g_{\mu \nu}$, considering two types of strings with 
different tensions. We call $g_{\mu \nu}$ the universal metric, which in fact does not necessarily satisfy Einstein´s equations.

 In the case of the flat space for the string associated metrics, in the Milne representation, for the case of two types of string tensions, we study the case where the two types of strings have positive string tensions , as opposed to our previous work \cite{cosmologyandwarped} where we found solutions with both positive and negative string tensions. At the early universe the negative string tension strings tensions are large in magnitude , but approach zero in the late universe and the  positive string tensions appear   for the late universe with their tension approaching a constant value at the late universe.  These solutions are absolutely singularity free.

In contrast, we have studied also \cite{Escaping} the case of very different solutions and where both
type of strings have positive tensions,  then these are singular, they cannot be continued before a certain time (that corresponded to a bounce in our previous work \cite{cosmologyandwarped}). Here, at the origin of time, the string tensions of both types of strings approach plus infinity, so this opens the possibility of having no Hagedorn temperature  in the early universe and latter on in the history of the universe as well for this type of string cosmology scenario. Also the tensions can become infinity at a certain location in the warped coordinate in a warped scenario.

Here we will study a situation where we consider the metrics $(\phi + T _{i})g_{\mu \nu}$, considering two types of strings tensions . The two metrics will again satisfy Einstein´s equations and the two metrics will represent Minkowski space and Minkowski space after a special conformal transformation. In this case, the location where the two types of strings acquire an infinite tension is given bu two surfaces. If the vector that defines the special conformal transformation is light like, these two surface are plane, parallel to each other and both move with the speed of light. If the vector is time like of space like the two surfaces are spherical and expanding and the distance between them approaches zero at large times (positive or negative) . In both cases this represents a genuine brane world scenario.

\section{The Modified Measure Theory String Theory}

The standard world sheet string sigma-model action using a world sheet metric is \cite{pol1}, \cite{pol2}, \cite{pol3}

\begin{equation}\label{eq:1}
S_{sigma-model} = -T\int d^2 \sigma \frac12 \sqrt{-\gamma} \gamma^{ab} \partial_a X^{\mu} \partial_b X^{\nu} g_{\mu \nu}.
\end{equation}

Here $\gamma^{ab}$ is the intrinsic Riemannian metric on the 2-dimensional string worldsheet and $\gamma = det(\gamma_{ab})$; $g_{\mu \nu}$ denotes the Riemannian metric on the embedding spacetime. $T$ is a string tension, a dimension full scale introduced into the theory by hand. \\

Now instead of using the measure $\sqrt{-\gamma}$ ,  on the 2-dimensional world-sheet, in the framework of this theory two additional worldsheet scalar fields $\varphi^i (i=1,2)$ are considered. A new measure density is introduced:

\begin{equation}
\Phi(\varphi) = \frac12 \epsilon_{ij}\epsilon^{ab} \partial_a \varphi^i \partial_b \varphi^j.
\end{equation}

There are no limitations on employing any other measure of integration different than $\sqrt{-\gamma}$. The only restriction is that it must be a density under arbitrary diffeomorphisms (reparametrizations) on the underlying spacetime manifold. The modified-measure theory is an example of such a theory. \\

Then the modified bosonic string action is (as formulated first in \cite{a} and latter discussed and generalized also in \cite{c})

\begin{equation} \label{eq:5}
S = -\int d^2 \sigma \Phi(\varphi)(\frac12 \gamma^{ab} \partial_a X^{\mu} \partial_b X^{\nu} g_{\mu\nu} - \frac{\epsilon^{ab}}{2\sqrt{-\gamma}}F_{ab}(A)),
\end{equation}

where $F_{ab}$ is the field-strength  of an auxiliary Abelian gauge field $A_a$: $F_{ab} = \partial_a A_b - \partial_b A_a$. \\

It is important to notice that the action (\ref{eq:5}) is invariant under conformal transformations of the internal metric combined with a diffeomorphism of the measure fields, 

\begin{equation} \label{conformal}
\gamma_{ab} \rightarrow J\gamma_{ab}, 
\end{equation}

\begin{equation} \label{diffeo} 
\varphi^i \rightarrow \varphi^{'i}= \varphi^{'i}(\varphi^i)
\end{equation}
such that 
\begin{equation} \label{measure diffeo} 
\Phi \rightarrow \Phi^{'}= J \Phi
\end{equation}

Here $J$ is the jacobian of the diffeomorphim in the internal measure fields which can be an arbitrary function of the world sheet space time coordinates, so this can called indeed a local conformal symmetry.

To check that the new action is consistent with the sigma-model one, let us derive the equations of motion of the action (\ref{eq:5}). \\

The variation with respect to $\varphi^i$ leads to the following equations of motion:

\begin{equation} \label{eq:6}
\epsilon^{ab} \partial_b \varphi^i \partial_a (\gamma^{cd} \partial_c X^{\mu} \partial_d X^{\nu} g_{\mu\nu} - \frac{\epsilon^{cd}}{\sqrt{-\gamma}}F_{cd}) = 0.
\end{equation}

since $det(\epsilon^{ab} \partial_b \varphi^i )= \Phi$, assuming a non degenerate case ($\Phi \neq 0$), we obtain, 

\begin{equation} \label{eq:a}
\gamma^{cd} \partial_c X^{\mu} \partial_d X^{\nu} g_{\mu\nu} - \frac{\epsilon^{cd}}{\sqrt{-\gamma}}F_{cd} = M = const.
\end{equation}

The equations of motion with respect to $\gamma^{ab}$ are

\begin{equation} \label{eq:8}
T_{ab} = \partial_a X^{\mu} \partial_b X^{\nu} g_{\mu\nu} - \frac12 \gamma_{ab} \frac{\epsilon^{cd}}{\sqrt{-\gamma}}F_{cd}=0.
\end{equation}

One can see that these equations are the same as in the sigma-model formulation . Taking the trace of (\ref{eq:8}) we get that $M = 0$. By solving $\frac{\epsilon^{cd}}{\sqrt{-\gamma}}F_{cd}$ from (\ref{eq:a}) (with $M = 0$) we obtain the standard string eqs. \\

The emergence of the string tension is obtained by varying the action with respect to $A_a$:

\begin{equation}
\epsilon^{ab} \partial_b (\frac{\Phi(\varphi)}{\sqrt{-\gamma}}) = 0.
\end{equation}

Then by integrating and comparing it with the standard action it is seen that

\begin{equation}
\frac{\Phi(\varphi)}{\sqrt{-\gamma}} = T.
\end{equation}

That is how the string tension $T$ is derived as a world sheet constant of integration opposite to the standard equation (\ref{eq:1}) where the tension is put ad hoc.
Let us stress that the modified measure string theory action 
does not have any \textsl{ad hoc} fundamental scale parameters. associated with it. This can be generalized to incorporate super symmetry, see for example \cite{c}, \cite{cnish}, \cite{supermod} , \cite{T1}.
For other mechanisms for dynamical string tension generation from added string world sheet fields, see for example \cite{xx} and \cite{xxx}. However the fact that this string tension generation is a world sheet effect 
and not a universal uniform string tension generation effect for all strings has not been sufficiently emphasized before.

Notice that Each String  in its own world sheet determines its own  tension. Therefore the  tension is not universal for all strings.

\section{Introducing  Background Fields including a New Background Field, The Tension Field} 
Schwinger \cite{Schwinger} had an important insight and understood that all the information concerning a field theory can be studied by understanding how it  reacts to sources of different types. 

This has been discussed in the text book by Polchinski for example  \cite{Polchinski} .  Then the target space metric and other  external fields acquire dynamics which is enforced by the requirement of zero beta functions.

However, in addition to the traditional background fields usually considered in conventional string theory, one may consider as well an additional scalar field that induces currents in the string world sheet and since the current couples to the world sheet gauge fields, this produces a dynamical tension controlled by the external scalar field as shown at the classical level in \cite{Ansoldi}. In the next two subsections we will study how this comes about in two steps, first we introduce world sheet currents that couple to the internal gauge fields in Strings and Branes and second we define a coupling to an external scalar field by defining a world sheet currents that couple to the internal gauge fields in Strings  that is induced by such external scalar field.

\subsection{Introducing world sheet currents that couple to the internal gauge fields}

If to the action of the string  we add a coupling
to a world-sheet current $j ^{a}$,  i.e. a term
\begin{equation}
    S _{\mathrm{current}}
    =
    \int d ^{2} \sigma
        A _{a}
        j ^{a}
    ,
\label{eq:bracuract}
\end{equation}
 then the variation of the total action with respect to $A _{a }$
gives
\begin{equation}
    \epsilon ^{a b}
    \partial _{a }
    \left(
        \frac{\Phi}{\sqrt{- \gamma}}
    \right)
    =
    j ^{b}
    .
\label{eq:gauvarbracurmodtotact}
\end{equation}
We thus see indeed that, in this case, the dynamical character of the
brane is crucial here.
\subsection{How a world sheet current can naturally be induced by a bulk scalar field, the Tension Field}

Suppose that we have an external scalar field $\phi (x ^{\mu})$
defined in the bulk. From this field we can define the induced
conserved world-sheet current
\begin{equation}
    j ^{b}
    =
    e \partial _{\mu} \phi
    \frac{\partial X ^{\mu}}{\partial \sigma ^{a}}
    \epsilon ^{a b}
    \equiv
    e \partial _{a} \phi
    \epsilon ^{a b}
    ,
\label{eq:curfroscafie}
\end{equation}
where $e$ is some coupling constant. The interaction of this current with the world sheet gauge field  is also invariant under local gauge transformations in the world sheet of the gauge fields
 $A _{a} \rightarrow A _{a} + \partial_{a}\lambda $.

For this case,  (\ref{eq:gauvarbracurmodtotact}) can be integrated to obtain
\begin{equation}
  T =  \frac{\Phi}{\sqrt{- \gamma}}
    =
    e \phi + T _{i}
    ,
\label{eq:solgauvarbracurmodtotact2}
\end{equation}
or  equivalently
\begin{equation}
  \Phi
    =
   \sqrt{- \gamma}( e \phi + T _{i})
    ,
\label{eq:solgauvarbracurmodtotact}
\end{equation}

The constant of integration $T _{i}$ may vary from one string to the other. Notice tha the interaction is metric independent since the internal gauge field does not transform under the the conformal transformations. This interaction does not therefore spoil the world sheet conformal transformation invariance in the case the field $\phi$ does not transform under this transformation.  One may interpret 
(\ref{eq:solgauvarbracurmodtotact} ) as the result of integrating out classically (through integration of equations of motion) or quantum mechanically (by functional integration of the internal gauge field, respecting the boundary condition that characterizes the constant of integration  $T _{i}$ for a given string ). Then replacing 
$ \Phi
    =
   \sqrt{- \gamma}( e \phi + T _{i})$ back into the remaining terms in the action gives a correct effective action for each string. Each string is going to be quantized with each one having a different $ T _{i}$. The consequences of an independent quantization of  many strings with different $ T _{i}$
covering the same region of space time will be studied in the next section.    

\subsection{Consequences from World Sheet  Quantum Conformal Invariance on the Tension field, when several strings share the same region of space}

\subsubsection{The case where all all string tensions are the same, i.e.,  $T _{i}=  T _{0}$, and the appearance of a target space conformal invariance }
If  all  $T _{i}=  T _{0}$, we just redefine our background field so that $e\phi+T _{0} \rightarrow  e\phi$ and then in the effective action for all the strings the same combination $e\phi g_{\mu \nu}$, 
and only this combination will be determined by the requirement that the conformal invariance in the world sheet of all strings be preserved quantum mechanically, that is , that the beta function be zero. So in this case we will not be able to determine $e\phi$ and 
$ g_{\mu \nu}$ separately, just the product $e\phi g_{\mu \nu}$,
so the equation obtained from equating the beta function to zero will have the target space conformal invariance 
$e\phi \rightarrow F(x)e\phi $, 
$g_{\mu \nu} \rightarrow F(x)^{-1}g_{\mu \nu} $. 

That is, there is no independent dynamics for the Tension Field in this case.
On the other hand, if there are at least two types of string tensions, that symmetry will not exist and there is the possibility of determining separately  $e\phi$ and 
$ g_{\mu \nu}$ as we will see in the next subsection.

\subsubsection{The case of two different string tensions }

If we have a scalar field coupled to a string or a brane in the way described in the sub section above, i.e. through the current induced by the scalar field in the extended object,  according to eq. 
(\ref{eq:solgauvarbracurmodtotact}), so we have two sources for the variability of the tension when going from one string to the other: one is the integration constant $T _{i}$ which varies from string to string and the other the local value of the scalar field, which produces also variations of the  tension even within the string or brane world sheet.

As we discussed in the previous section, we can incorporate the result of the tension as a function of scalar field $\phi$, given as $e\phi+T_i$, for a string with the constant of integration $T_i$ by defining the action that produces the correct 
equations of motion for such string, adding also other background fields, the anti symmetric  two index field $A_{\mu \nu}$ that couples to $\epsilon^{ab}\partial_a X^{\mu} \partial_b X^{\nu}$
and the dilaton field $\varphi $ that couples to the topological density $\sqrt{-\gamma} R$
\begin{equation}\label{variablestringtensioneffectiveacton}
S_{i} = -\int d^2 \sigma (e\phi+T_i)\frac12 \sqrt{-\gamma} \gamma^{ab} \partial_a X^{\mu} \partial_b X^{\nu} g_{\mu \nu} + \int d^2 \sigma A_{\mu \nu}\epsilon^{ab}\partial_a X^{\mu} \partial_b X^{\nu}+\int d^2 \sigma \sqrt{-\gamma}\varphi R .
\end{equation}
Notice that if we had just one string, or if all strings will have the same constant of integration $T_i = T_0$.

In any case, it is not our purpose here to do a full generic analysis of all possible background metrics, antisymmetric two index tensor field and dilaton fields, instead, we will take  cases where the dilaton field is a constant or zero, and the antisymmetric two index tensor field is pure gauge or zero, then the demand of conformal invariance for $D=26$ becomes the demand that all the metrics
\begin{equation}\label{tensiondependentmetrics}
g^i_{\mu \nu} =  (e\phi+T_i)g_{\mu \nu}
\end{equation}
will satisfy simultaneously the vacuum Einstein´s equations,
Notice that if we had just one string, or if all strings will have the same constant of integration $T_i = T_0$, then all the 
$g^i_{\mu \nu}$ metrics are the same and then 
(\ref{tensiondependentmetrics}) is just a single field redefinition and therefore there will be only one metric that will have to satisfy Einstein´s equations, which of course will not impose a constraint on the tension field $\phi$ . 

The interesting case to consider is therefore many strings with different $T_i$, let us consider the simplest case of two strings, labeled $1$ and $2$ with  $T_1 \neq  T_2$ , then we will have two Einstein´s equations, for $g^1_{\mu \nu} =  (e\phi+T_1)g_{\mu \nu}$ and for  $g^2_{\mu \nu} =  (e\phi+T_2)g_{\mu \nu}$, 

\begin{equation}\label{Einstein1}
R_{\mu \nu} (g^1_{\alpha \beta}) = 0 
\end{equation}
and , at the same time,
\begin{equation}\label{Einstein1}
  R_{\mu \nu} (g^2_{\alpha \beta}) = 0
\end{equation}

These two simultaneous conditions above impose a constraint on the tension field
 $\phi$, because the metrics $g^1_{\alpha \beta}$ and $g^2_{\alpha \beta}$ are conformally related, but Einstein´s equations are not conformally invariant, so the condition that Einstein´s equations hold  for both  $g^1_{\alpha \beta}$ and $g^2_{\alpha \beta}$
is highly non trivial.

Let us consider the case that one of the metrics, say  $g^2_{\alpha \beta}$ is a Schwarzschild solution, either a 
4 D Schwarzschild solution X a product flat of Torus compactified extra dimensions
or just a 26 D Schwarzschild solution, in this case, it does not appear possible to have a conformally transformed  $g^2_{\alpha \beta}$ for anything else than in the case that the conformal factor that transforms the two metrics is a positive constant, let us call it $\Omega^2$, in that case $g^1_{\alpha \beta}$ is a Schwarzschild solution of the same type, just with a different mass parameter and different sizes of extra dimensions if the compactified solution is considered. Similar consideration holds for the case the 2 metric is a Kasner solution,

Then  in this case also, it does not appear possible to have a conformally transformed  $g^2_{\alpha \beta}$ for anything else than in the case that the conformal factor that transforms the two metrics is a constant, we will find other cases where the conformal factor will not be a constant,  let us call then conformal factor $\Omega^2$ in general, even when it is not a constant.
One can also study metrics used to describe gravitational radiation,
then again, multiplying by a constant both the background flat space and the perturbation gives us also a solution of vacuum Einstein´s equations. 

Then for these situations, we have,
\begin{equation}\label{relationbetweentensions}
e\phi+T_1 = \Omega^2(e\phi+T_2)
\end{equation}
 which leads to a solution for $e\phi$
 
\begin{equation}\label{solutionforphi}
e\phi  = \frac{\Omega^2T_2 -T_1}{1 - \Omega^2} 
\end{equation}
which leads to the tensions of the different strings to be
\begin{equation}\label{stringtension1}
 e\phi+T_1 = \frac{\Omega^2(T_2 -T_1)}{1 - \Omega^2} 
\end{equation}
and
  \begin{equation}\label{stringtension2}
 e\phi+T_2 = \frac{(T_2 -T_1)}{1 - \Omega^2} 
\end{equation}

Both tensions can be taken as positive if $T_2 -T_1$ is positive and $\Omega^2$ is also positive and less than $1$.
It is important that we were forced to consider a multi metric situation. One must also realize that  $\Omega^2$ is physical, 
because both metrics live in the same spacetime, so even if $\Omega^2$ is a constant ,  we are not allowed to perform a coordinate transformation, consisting for example of a rescaling of coordinates for one of the metrics and not do the same transformation for the other metric. 

Other way  to see that $\Omega^2$ is physical consist of considering the scalar consisting of the ratio of the two measures $\sqrt{-g^1}$ and $\sqrt{-g^2}$ where $ g^1 = det ( g^1_{\alpha \beta})$ and $ g^2 = det ( g^2_{\alpha \beta})$, and we find that the scalar 
$\frac{\sqrt{-g^1}}{\sqrt{-g^2}} = \Omega^{D}$, showing that $\Omega$ is a coordinate invariant. 

\subsubsection{Flat space in Minkowski coordinates and flat space after a special conformal transformation }

Let us study now a case where $\Omega^2$ is not a constant. For this we will consider two spaces related by a conformal transformation, which will be flat space in Minkowski coordonates and flat space after a special conformal transformation. 
 
The flat space in Minkowskii coordinates is,

 \begin{equation}\label{Minkowski}
 ds_1^2 = \eta_{\alpha \beta} dx^{\alpha} dx^{\beta}
\end{equation}

where $ \eta_{\alpha \beta}$ is the standard Minkowski metric, with 
$ \eta_{00}= 1$, $ \eta_{0i}= 0 $ and $ \eta_{ij}= - \delta_{ij}$.
This is of course a solution of the vacuum Einstein´s equations.

We now consider the conformally transformed metric

 \begin{equation}\label{Conformally transformed Minkowski}
 ds_2^2 = \Omega(x)^2  \eta_{\alpha \beta} dx^{\alpha} dx^{\beta}
\end{equation}
which we also demand that will satisfy the $D$ dimensional  vacuum Einstein´s equations.

Let us use the known transformation law of the Ricci tensor under a conformal transformation applied to 
$g^1_{\alpha \beta}=\eta_{\alpha \beta}$ and 
$  g^2_{\alpha \beta}=\Omega(x)^2\eta_{\alpha \beta}$,
defining $ \Omega(x)=  \theta^{-1}$, we obtain
\begin{equation}\label{Conformally transformed Ricci}
\begin{split}
R^2_{\alpha \beta}=& R^1_{\alpha \beta} + (D-2)
\nabla_{\alpha}\nabla_{\beta}(ln \theta) + \eta _{\alpha \beta}\eta^{\mu \nu}\nabla_{\mu}\nabla_{\nu}(ln \theta)+(D-2)\nabla_{\alpha}(ln \theta)\nabla_{\beta}(ln \theta) \\
& -(D-2)\eta _{\alpha \beta}\eta^{\mu \nu}\nabla_{\mu}(ln \theta)\nabla_{\nu}(ln \theta)
\end{split}
\end{equation}

Since $g^1_{\alpha \beta}=\eta_{\alpha \beta}$, we obtain that $R^1_{\alpha \beta} =0$, also the covariant derivative above are covariant derivatives with respect to the metric  $g^1_{\alpha \beta}=\eta_{\alpha \beta}$, so they are just ordinary derivatives. Taking this into account, after a bit of algebra we get that,
\begin{equation}\label{Conformally transformed Ricci}
\begin{split}
R^2_{\alpha \beta}=&  (D-2)
\frac{\partial_{\alpha}\partial_\beta \theta}{\theta}+ \eta _{\alpha \beta}\eta^{\mu \nu}(\frac{\partial_{\mu}\partial_{\nu}\theta}{\theta}-\frac{\partial_{\mu}\theta\partial_{\nu}\theta}{\theta^2})\\
& -(D-2)\eta _{\alpha \beta}\eta^{\mu \nu}\frac{\partial_{\mu}\theta\partial_{\nu}\theta}{\theta^2} = 0
\end{split}
\end{equation}
by contracting (\ref{Conformally transformed Ricci}) we obtain a relation between $\eta^{\mu \nu}\frac{\partial_{\mu}\partial_{\nu}\theta}{\theta}$ and  $\eta^{\mu \nu}{\partial_{\mu}\theta\partial_{\nu}\theta}/\theta^2$

\begin{equation}\label{relation}
2\eta^{\mu \nu}\frac{\partial_{\mu}\partial_{\nu}\theta}{\theta} =
D\frac{\eta^{\mu \nu}{\partial_{\mu}\theta\partial_{\nu}\theta}}{\theta^2}
\end{equation}
using (\ref{relation}) to eliminate the nonlinear term $\eta^{\mu \nu}\frac{\partial_{\mu}\theta\partial_{\nu}\theta}{\theta}$ in (\ref{Conformally transformed Ricci}) we obtain the remarkably simple linear relation,
 \begin{equation}\label{linear relation}
 \partial_{\alpha}\partial_\beta \theta -
 \frac{1}{D}\eta_{\alpha \beta}\eta^{\mu \nu}\partial_{\mu}\partial_{\nu}\theta = 0
\end{equation}
 
 So we now first find the most general solution of the linear equation
 (\ref{linear relation}),  which is,
 \begin{equation}\label{solution of linear relation}
 \theta= a_1 + a_2 K_{\mu}x^{\mu} +  a_3 x^{\mu}x_{\mu}
 \end{equation}
 and then impose the non linear constraint (\ref{relation}), which implies, 
 \begin{equation}\label{solution of a1}
 a_1 = \frac{a^2_2 K_{\mu} K^{\mu}}{4  a_3}
 \end{equation}
 
 we further demand that $\theta(x^{\mu}=0)=1$, so that,
 \begin{equation}\label{solution of linear relation}
 \theta= 1 + a_2 K_{\mu}x^{\mu} +  \frac{a^2_2 K_{\mu} K^{\mu}}{4} x^{\mu}x_{\mu}
 \end{equation}
 This coincides with the results of Culetu \cite{Culetu} for $D=4$
 and to identify this result with the result of a special conformal transformation, see discussions in  \cite{Kastrup} and \cite{Zumino} , to connect to standard notation we identify $a_2 K_{\mu} = 2 a_{\mu}$, so that
  \begin{equation}\label{conformal factor of special conformal transformation}
 \theta= 1 +2 a_{\mu}x^{\mu} +   a^2 x^2
 \end{equation}
 where $ a^2 =a^{\mu}a_{\mu}$ and $ x^2= x^{\mu}x_{\mu}$. 

In this case, this conformal factor coincides with that obtained from the special conformal transformation
\begin{equation}\label{ special conformal transformation}
x\prime ^{\mu} =  \frac{(x ^{\mu} +a ^{\mu} x^2)}{(1 +2 a_{\nu}x^{\nu} +   a^2 x^2)}
 \end{equation}
 
 As discussed by Zumino \cite{Zumino} the finite special conformal transformation mixes up in a complicated way the topology of space time, so it is not useful to interpret the finite special conformal transformations as mapping of spacetimes.
 
 In summary, we have two solutions for the Einstein´s equations,
 $g^1_{\alpha \beta}=\eta_{\alpha \beta}$ and 
 
 \begin{equation}\label{ conformally transformed metric}
 g^2_{\alpha \beta}= \Omega^2\eta_{\alpha \beta} =\theta^{-2}\eta_{\alpha \beta}=\frac{1}{( 1 +2 a_{\mu}x^{\mu} +   a^2 x^2)^2} \eta_{\alpha \beta}
 \end{equation}
 
 We can then study the evolution of the tensions using 
 $\Omega^2 =\theta^{-2}=\frac{1}{( 1 +2 a_{\mu}x^{\mu} +  a^2 x^2)^2}$.
 We will consider two different cases: 1) $ a^2 =0 $, 2)  $a^2 \neq 0 $
 \subsubsection{Light Like Segment Compactification }
 Here we consider the case  $ a^2 =0 $, and let us consider 
  $ a^{\mu} = (A,A,0,......0) $. Then 
 \begin{equation}\label{ conformally factor shock wave}
\Omega^2 =\frac{1}{( 1 +2 a_{\mu}x^{\mu})^2}=\frac{1}{( 1 +2 A(t-x))^2}
 \end{equation}
 
 From this, let is calculate the tensions of the two sting types and see that they will be constrained to be inside a segment that moves with the speed of light. At the boundaries of those segments the string tensions become infinity, so the strings cannot escape this segment.
 
\ref{ conformally factor shock wave} leads to the tensions of the different strings to be
\begin{equation}\label{stringtension1segment}
 e\phi+T_1 = \frac{\Omega^2(T_2 -T_1)}{1 - \Omega^2} =
 \frac{(T_2 -T_1)(1 +2A(t-x))^2}{4A(t-x)(1+A(t-x))}
\end{equation}
and
  \begin{equation}\label{stringtension2segment}
 e\phi+T_2 = \frac{(T_2 -T_1)}{1 - \Omega^2} = 
 \frac{(T_2 -T_1)}{4A(t-x)(1+A(t-x))}
\end{equation}
Let us take $T_2 -T_1$ positive, A negative, so we see that both tensions above go to positive infinity when $t-x$ goes to zero from negative values . Also  both tensions above go to positive infinity when $t-x$ goes to the value $-1/A$ from above.
That means that the strings are confined to the moving segment where 
  $t-x$ is inside the segment $(-1/A ,0 )$. We call this phenomenon ¨Light Like Segment Compactification¨. 
  Light like extra dimensions have been considered by Searight  
   \cite{Searight} and Braneworlds via Lightlike Branes was considered in \cite{Braneworlds via Lightlike Branes} for example.
   
   To complete the discussion of the Light Like Segment Compactification, we present the universal metric. From the relation $g^1_{\mu \nu} =  (e\phi+T_1)g_{\mu \nu}$ and the solution for $(e\phi+T_1)$ from \ref{stringtension1segment}, we obtain
   \begin{equation}\label{universal metric light like}
g_{\mu \nu}  =  \frac{1}{(e\phi+T_1)} g^1_{\mu \nu} = 
 \frac{4A(t-x)(1+A(t-x))}{(T_2 -T_1)(1 +2A(t-x))^2}\eta_{\mu\nu}
\end{equation}
   showing  this metric goes through zero changes sign at the boundaries of the segment.
   By considering the strings confined to the segment we avoid this pathology, this provides another way to justify that we must have a light like segment compactication.
   In the next subsection we will see how thick, expanding at sub luminal velocities braneworld
   scenarios are obtained from Dynamical  String  Tension Theories.
   
  \subsubsection{ Braneworlds  in Dynamical  String  Tension Theories }
  We now consider the case when $a^\mu$ is not light like and we will find that for $a^2 \neq 0$, irrespective of sign, i.e. irrespective of whether  $a^\mu$ is space like or time like, we will have thick  Braneworlds  where strings can be constrained  between two concentric spherically symmetric bouncing higher dimensional spheres and where the distance between these two  concentric spherically symmetric bouncing higher dimensional spheres approaches zero at large times.
  The string tensions of the strings one and two are given by
    \begin{equation}\label{stringtension1forBraneworld}
 e\phi+T_1 = \frac{(T_2-T_1)( 1 +2 a_{\mu}x^{\mu} +  a^2 x^2)^2}{( 1 +2 a_{\mu}x^{\mu} +  a^2 x^2)^2-1}=  \frac{(T_2-T_1)( 1 +2 a_{\mu}x^{\mu} +  a^2 x^2)^2}{(2 a_{\mu}x^{\mu} +  a^2 x^2)(2+2 a_{\mu}x^{\mu} +  a^2 x^2)}
\end{equation}
  \begin{equation}\label{stringtension2forBraneworld}
 e\phi+T_2 = \frac{(T_2-T_1)}{( 1 +2 a_{\mu}x^{\mu} +  a^2 x^2)^2-1}=  \frac{(T_2-T_1)}{(2 a_{\mu}x^{\mu} +  a^2 x^2)(2+2 a_{\mu}x^{\mu} +  a^2 x^2)}
\end{equation}
  
Then, the locations where the string tensions go to infinity are determined by the conditions

\begin{equation}\label{boundariesforBraneworld1}
2 a_{\mu}x^{\mu} +  a^2 x^2 = 0
\end{equation}
or 
\begin{equation}\label{boundariesforBraneworld2}
2 +2 a_{\mu}x^{\mu} +  a^2 x^2 = 0
\end{equation}
Let us start by considering the case where  $a^\mu$ is time like, then without loosing generality we can take  $a^\mu = (A, 0, 0,...,0)$.
In this case the denominator in (\ref{stringtension1forBraneworld}) , (\ref{stringtension2forBraneworld}) is
\begin{equation}\label{denominatortimelike}
(2 a_{\mu}x^{\mu} +  a^2 x^2)(2+2 a_{\mu}x^{\mu} +  a^2 x^2) =
(2At +A^2(t^2-x^2))(2+2At++A^2(t^2-x^2))
\end{equation}

The condition (\ref{boundariesforBraneworld1}) implies then that
\begin{equation}\label{bubbleboundaryforBraneworld1a}
 x^2_1  + x^2_2 + x^2_3.....+ x^2_{D-1}- (t+ \frac{1}{A})^2 = -\frac{1}{A^2}
\end{equation}
while the other boundary of infinite string tension (\ref{boundariesforBraneworld2}) is given by,
\begin{equation}\label{bubbleboundaryforBraneworld1b}
 x^2_1  + x^2_2 + x^2_3.....+ x^2_{D-1}- (t+ \frac{1}{A})^2 = \frac{1}{A^2}
\end{equation}
So we see that (\ref{bubbleboundaryforBraneworld1b}) represents an exterior boundary which has an bouncing  motion with a minimum radius $\frac{1}{A}$ at $t = - \frac{1}{A}$ , 
The denominator (\ref{denominator}) is positive between these two bubbles.
So for $T_2 -T_1$ positive the tensions are positive and diverge at the boundaries defined above.

The internal boundary (\ref{bubbleboundaryforBraneworld1a}) exists only for times $t$ smaller than $-\frac{2}{A}$ and bigger than  
$0$, so in the time interval $(-\frac{2}{A},0)$
there is no inner surface of infinite tension strings.
This inner surface collapses to zero radius at  $t=-\frac{2}{A}$
and emerges again from zero radius at $t=0$.

For large positive or negative times, the difference between the upper radius  and the lower radius goes to zero as  $t \rightarrow \infty$

\begin{equation}\label{asymptotic}
\sqrt{\frac{1}{A^2} +(t+ \frac{1}{A})^2 } -\sqrt{-\frac{1}{A^2} +(t+ \frac{1}{A})^2 }\rightarrow \frac{1}{t A^2}\rightarrow 0  
\end{equation}
of course the same holds  $t \rightarrow -\infty$.
This means that for very large early or late times the segment where the strings would be confined (since they will avoid having infinite tension) will be very narrow and the resulting scenario will be that od a brane world for late or early times, while in the bouncing region the inner surface does not exist.
Let us start by considering the case where  $a^\mu$ is space like, then without loosing generality we can take  $a^\mu = (0, A, 0,...,0)$.
In this case the denominator in (\ref{stringtension1forBraneworld}) , (\ref{stringtension2forBraneworld}) is
\begin{equation}\label{denominatorspacelike}
(2 a_{\mu}x^{\mu} +  a^2 x^2)(2+2 a_{\mu}x^{\mu} +  a^2 x^2) =
(-2Ax^1-A^2(t^2- \vec{x}^2))((2-2Ax^1-A^2(t^2-\vec{x}^2))
\end{equation}
where $\vec{x}= (x^1, x^2,...., x^{D-1})$ represents the spacial part  of $x^{\mu}$, and  $\vec{x}^2= (x^1)^2 +(x^2)^2+....+ (x^{D-1})^2$.
We now consider the case when   $a^\mu$ is space like, then without loosing generality we can take  $a^\mu = (0, A , 0,...,0)$.
We then consider the first boundary where the string tensions approoach infinity according to (\ref{boundariesforBraneworld1}),

\begin{equation}\label{bubbleboundaryforBraneworld1b}
-( x_1 -\frac{1}{A})^2  - x^2_2 - x^2_3.....- x^2_{D-1}+ t^2 = -\frac{1}{A^2}
\end{equation}
which describes a bouncing bubble with minimum radius $\frac{1}{A}$
at $t=0$.

The case (\ref{boundariesforBraneworld2}) gives
\begin{equation}\label{bubbleboundaryforBraneworld2b}
-( x_1 -\frac{1}{A})^2  - x^2_2 - x^2_3.....- x^2_{D-1}+ t^2 = \frac{1}{A^2}
\end{equation}
(\ref{bubbleboundaryforBraneworld2b}) is an  internal boundary which  exists only for times $t$ smaller than $-\frac{1}{A}$ and bigger than  
$\frac{1}{A}$. Between  $-\frac{1}{A}$ and   
$\frac{1}{A}$ there is no inner surface of infinite tension strings.
Between these two bubbles the two factors in eq. 
This inner surface collapses to zero radius at  $t=-\frac{1}{A}$
and emerges again from zero radius at $t=-\frac{1}{A}$. 
So the situation is very similar to that of the case where the vector 
 $a^\mu$ is time like, just that the roles of the cases $\Omega = 1$ and $\Omega = -1$ get exchanged. 
 Between these two boundaries the two factors in the denominator   (\ref{denominatorspacelike}) are positive, while at the boundaries one or the other approach zero and the tensions diverge, so again for $T_2 -T_1$ positive the tensions are positive and diverge at the boundaries.
 
 Once again for large positive or negative times, the difference between the upper radius  and the lower radius goes to zero. Implying 
 that the strings will be confined to a very small segment at large early or late times, so then again we get an emergent brane world scenario.
 
 The strings and therefore all matter and gravity will be consequently confined to the very small segment of size $ \frac{1}{t A^2}$, very small for large $t$. At the moment of the bounce there is no brane world, there is only one exterior bubble which represents infinite tension location, the brane is generated dynamically after a period of time by the appearance of the inner bubble which completes the trapping of the strings between two surfaces.

   To complete the discussion of the braneworld case, we present the universal metric. From the relation $g^1_{\mu \nu} =  (e\phi+T_1)g_{\mu \nu}$ and the solution for $(e\phi+T_1)$ from (\ref{stringtension1segment}), we obtain,
   \begin{equation}\label{universal metric braneworld}
g_{\mu \nu}  =  \frac{1}{(e\phi+T_1)} g^1_{\mu \nu} = 
 \frac {(2 a_{\mu}x^{\mu} +  a^2 x^2)(2+2 a_{\mu}x^{\mu} +  a^2 x^2)}{(T_2-T_1)( 1 +2 a_{\mu}x^{\mu} +  a^2 x^2)^2}\eta_{\mu\nu}
\end{equation}
   showing that this metric becomes zero at the boundaries where the tensions go infinity and if we extend the metric beyond the boundary the metric changes signature. If our basic postulate is that the signature of the metric $g_{\mu \nu}$ is $(+,-,-,-,-,-,-,-,....)$, then we cannot allowed to extend the space time beyond the region of space time where the string tensions are positive.
   By considering the strings confined to the boundaries we have defined we avoid these pathologies, this provides another way to justify that we must have a braneworld scenario.
 \section{Discussion: Comparison with Standard Approaches to Braneworlds and Perspectives}
 Our approach is very different to the present standard approaches to braneworlds in the context of string theories however. In our approach a dynamical string tension has been used. Our scenario could be enriched by incorporating aspects of the more traditional braneworlds, like introducing D-branes between the surfaces where the string tensions go to infinity, so open strings could end before their tensions approach an infinite value, or the surfaces where the tensions diverge could be themselves be defined as D branes for open strings. These possibilities have been ignored here to simplify the discussion.

In any case, given that the tension of the strings diverge at the two boundaries we have defined, all strings are confined between those, the closed strings also, so unlike more traditional braneworlds, gravity does not escape to the bulk, in fact in the framework proposed here a braneworld scenario using just closed strings is perfectly possible.

In spite of the differences with the more conventional approaches, one should question nevertheless whether still some of the typical signatures of these approaches. 
To start with, in order to calculate some of the features of the brane,  one would have to see if the branes we have obtained here have some flexibility or if they are  rigid. The degree of flexibility of the brane is measured by associating a tension to the brane. 
For a small brane tension, a flexible brane-world model,  brane excitations, branons, will be relevant and in this case branons are the only new relevant low-energy particles \cite{Branons} .
So far our calculations do not give us an indication concerning the tension of these branes, if this tension is big, small or infinite,
since we have seen the brane appearing in a most symmetric way. In order to see if the brane is rigid, we would have to consider how the solution when perturbed responds, from this we could identify a brane tension. This is a project for further research. 
Another subject which is very important is the combination of the braneworld with  compactification of some dimensions. Indeed , the brane makes one dimension small and this would be enough if we ere considering just five dimensions but in string theory we must consider $26$ dimensions, so additional ways to reduce effective dimensions to four must be considered, compactification as in Kaluza Klein scenarios comes to mind. In this respect, it is important that the brane obtained here are thick $D$ dimensional with one dimension getting smaller and smaller as times increases, so effectively we obtain a $D-1$ dimensional brane moving in a $D$ dimensional  universe, as compared to a $4$ brane (usually denoted as a 3 brane) where SM particles live in our 4 dimensional manifold used in standard string theory brane worlds \cite{Antoniadis}. Then according to the standard scenario, some other fields like gravity will live in the bulk . Here there will be no bulk , all fields live in the initially thick brane, which as time advances becomes a very thin brane, thinner and thinner
as time advances and the calculation of the modes at asymptotic times should be calculated in a $D-1$ with  a number of these dimensions compactified, so that the remaining uncompactified space time coordinates is only four. So, for the braneworld scenario advocated here , we could proceed as in ref \cite{KK} , but with the crucial modification that gravity is not on the bulk but rather in the brane,
in fact nothing should be on the bulk.
In the early universe however the brane is thicker, so cavity excitations,  where field reflect from the boundaries of the thick brane , associated casimir effects could be important there. There could be also interactions between the cavity modes and the Kaluza Klein excitations that could modify the predictions for the Kaluza Klein gravitons which are obtained in the more conventional braneworld models \cite{KK}.

Finally, since the string tensions go to infinity at the boundaries, there is the option of avoiding the Hagedorn Temperature, which is proportional to the string tension,  as it was discussed in the case of other examples where  the string tensions go to infinity in certain regions \cite{Escaping}. This possibility to obtain very high temperatures  could also give rise to interesting observational consequences for the graviton spectrum. This could have interesting consequences for the early universe, or collision experiments where there will be the possibility of thermalization.
 
\textbf{Acknowledgments}
 I thank Oleg Andreev, David Andriot, Stefano Ansoldi, David Benisty , Thomas Curtright, Euro Spallucci, Emil Nissimov, Svetlana Pacheva, Tatiana Vulfs,  Hitoshi Nishino, Subhash Rajpoot, Luciano Rezzolla, Horst Stöcker, Jurgen Struckmeier , David Vasak, Johannes Kirsch, Dirk Kehm , Luca  Mezincescu and  Matthias Hanauske for usefull discussions. I also want to thank the Foundational Questions Institute (FQXi)  and the COST actions  Quantum Gravity Phenomenology in the multi messenger approach, CA18108 and  Gravitational waves, Black Holes and Fundamental Physics, CA16104 for support and special thanks to the Frankfurt Institute for Advanced Study for hospitality and finantial support and to Astrophysics Group at Goethe University for  providing me with the opportunity of presenting these results in a seminar at one of their Astro Coffe Seminars, 
 https://astro.uni-frankfurt.de/astrocoffee/ and to The University of Miami for the possibility of presenting these results at the Miami 2021 conference https://cgc.physics.miami.edu/Miami2021/.

\end{document}